# FAIR data – new horizons for materials research

Matthias Scheffler[1,2], Martin Aeschlimann[3], Martin Albrecht[4], Tristan Bereau[5], Hans-Joachim Bungartz[6], Claudia Felser[7], Mark Greiner[8], Axel Groß[9], Christoph T. Koch[2], Kurt Kremer[5], Wolfgang E. Nagel[10], Markus Scheidgen[2], Christof Wöll[11], and Claudia Draxl[2,1]

[1] Fritz-Haber-Institut der Max-Planck-Gesellschaft, Berlin
[2] Physics Department and IRIS Adlershof, Humboldt-Universität zu Berlin, Berlin
[3] Department of Physics and Research Center OPTIMAS, University of Kaiserslautern, Kaiserslautern
[4] Leibniz-Institut für Kristallzüchtung, Berlin
[5] Max Planck Institute for Polymer Research, Mainz
[6] Department of Informatics, Technical University Munich, Munich
[7] Max Planck Institute for Chemical Physics of Solids, Dresden
[8] Max Planck Institute for Chemical Energy Conversion, Mülheim a.d. Ruhr
[9] Institute of Theoretical Chemistry, Ulm University and Helmholtz-Institute Ulm, Ulm
[10] Computer Science Department, Technical University Dresden, Dresden
[11] Institute of Functional Interfaces, Karlsruhe Institute of Technology, Karlsruhe

Corresponding author: claudia.draxl@physik.hu-berlin.de

**Abstract / Preface**

Prosperity and lifestyle of our society are very much governed by achievements of condensed-matter physics, chemistry, and materials science because new products for the energy, environment, health, mobility, IT sectors, *etc.*, largely rely on improved or even novel materials. Examples are solid-state lighting, touch screens, batteries, implants, drug delivery, and many more. The enormous amounts of research data produced every day in this field represent a gold mine of the 21$^{st}$ century. This gold mine is, however, of little value, if these data are not comprehensively characterized and made available. How can we refine this feedstock, *i.e.*, turn data into knowledge and value? For this, a *FAIR* (*Findable*, *Accessible*, *Interoperable*, and *Reusable*) data infrastructure is a must. Only then, data can be readily shared and explored by data analytics and artificial-intelligence (AI) methods. Making data *Findable and AI Ready* (a forward-looking interpretation of the acronym) will change the way how science is done today. In this *Perspective*, we discuss how we prepare to make this happen for the field of materials science.

**Introduction**

The number of possible materials is practically infinite. But even for the so far known materials, our knowledge about their properties and their synthesis is very shallow. There is no doubt that forms of condensed matter exist, or can be created, that exhibit better, or even novel, properties and functions than the materials that are known and used today. How can we find them? High-throughput screening of materials – experimentally or theoretically – collects important information. These results will boost novel discoveries, but the immensity of possible materials cannot be covered by such explicit searches. Moreover, in the present purpose-focused research, only a small fraction of the data produced in the studies is published, and many data are not fully characterized. Furthermore, metadata[a], ontologies[b], and workflows of different research groups cannot be easily reconciled. Thus, most research data are neither findable nor interoperable.

A *FAIR* data infrastructure will foster the exchange of scientific information. The meaning of the acronym, *i.e.*, that data should be *Findable*, *Accessible*, *Interoperable*, and *Re-usable* is explained in the original publication by Wilkinson and coworkers[1] and elaborated on, for example, at the GoFAIR web pages[2]. The crucial and very laborious first step towards the FAIRification of data concerns the need to comprehensively describe data by metadata, *i.e.*, to characterize data fully and unambiguously so that the research is reproducible. Then scientists, engineers, and others are also able to combine data and metadata of different studies and to use them in different contexts. This will open synergies between materials-science sub-domains and facilitate inter-institute and cross-discipline research. It will also enable that data can be used for deeper analyses and training AI models. Clearly, a *FAIR* data infrastructure will also reveal data provenance.

The US Materials Genome (MGI) Initiative was announced in 2011 for "*discovering, manufacturing, and deploying advanced materials twice as fast and at a fraction of the cost compared to traditional methods*".[3] It significantly boosted collaborations and high-throughput experiments and computations. The consortium *FAIRmat*[4] develops the original MGI concept further. It aims at implementing a *FAIR* research-data infrastructure that interweaves data and tools from and for materials synthesis, experiment, theory and computation, and makes all data available to the whole materials-science community and beyond. In this endeavor, it unites researchers from condensed-matter physics, the chemical physics of solids, and computer science with IT experts.

---

[a] With metadata we mean the information that explains and characterizes the measured or calculated data.
[b] An ontology describes the relationships in metadata. In general, it can help to describe workflows and results, and to integrate them into a bigger context.

Materials science is strongly affected by all the *4V of Big Data*, which are *volume* (the amount of data), *variety* (the heterogeneity of form and meaning of data), *velocity* (the rate at which data may change or new data arrive), and *veracity* (the uncertainty of data quality). The various experimental and theoretical examples provided below will illustrate these different aspects. In general, a *FAIR* data infrastructure requires an in-depth description of how the data have been obtained, *i.e.*, *addressing metadata, ontologies, and workflows.* Obviously, only the experts, *i.e.*, those creating the samples or computer codes, and performing measurements or calculations (*i.e.*, producing the data), have the insight and knowledge for providing this critical information.

The topic of this perspective, as sketched above, includes the request for a significant change of the scientific culture. Thinking beyond our present effect-, phenomenon-, or application-focused research, requires accepting that "clean data", *i.e.*, well-characterized and annotated data, represent a value of similar importance as a standard publication, or even higher. This concept carries analogies to Tycho Brahe who created the data that enabled Johannes Kepler to find his equations and finally led Newton to formulate his theory of gravitation.

Eventually, after having installed an efficient, *FAIR* research-data infrastructure, hosting all data from synthetic, experimental, and theoretical studies for a wide range of materials, we also need to pave the way for doing novel research. Our scientific vision is to build *maps of material properties* that will guide us in designing and finding new materials for a desired function. This concept follows the spirit of the creation of the *Periodic Table of Elements*: Organizing the about 60 known table entries enabled Mendeleev to predict the existence and properties of yet to be discovered elements.

In the following, we will describe the state-of-the-art, highlight the challenges, and put forward *FAIRmat*'s envisaged solutions.

**Data-centric materials science**

Science is and always has been based on data, but the term *data-centric* indicates a radical shift in the way information is handled and research is performed. It refers to e*xtensive* data collections, *digital* repositories, and new concepts and methods of *data analytics*. It also implies that we complement the traditional purpose-oriented research by using data from other studies.

Some progress in this direction has been made in recent years to collect data from the many research groups across the planet (all the data, not just what is published in research manuscripts) and to make the data *FAIR*[1,5]. This should be good scientific practice, anyhow[6,7]. Since 1965, data repositories in materials science have moved towards digitalization. A comprehensive list can be found in Ref. 8. Among them, the *NOMAD (Novel Materials Discovery) Laboratory* (a database for computational materials science; online since 2014)[9], is unique as it accepts data from practically all computational materials-science codes. Since it provides the blueprint of *FAIRmat*, let us summarize its basic concept (for details see Refs. 5 and 10). A key guideline of *NOMAD* (and *FAIRmat*) is to help scientists and students to up- and download data in a most comfortable way. In simple words, data stored at *NOMAD* are treated analogously to publications at a journal archive, *e.g.*, Ref. 11. Different to journal archives, there is the possibility of an embargo period that can be used for collaborations with selected colleagues or may be even crucial for collaborations with industry. At this point (August 2021), *NOMAD* contains results from more than 100 million open-access calculations. These are from individual researchers from all over the world and include entries from other computational materials databases, *e.g.*, *AFLOW*[12], *The Materials Project*[13], and *OQMD*[14]. *NOMAD* converts the data into a common form, provides an easy materials-view presentation (*NOMAD Encyclopedia*[15]), and tools for data analytics and predictions (*NOMAD AI Toolkit*[16]).

The overall challenges of *FAIRmat* are sketched in Fig. 1: Besides organizing and –equally important – convincing the community (top left), a critical task concerns the development of metadata standards and ontologies (top right). Presently, in materials science, such standards are either totally missing or incomplete. Numerous attempts from standards organizations, such as the *International Standards Organization (ISO),* to provide controlled vocabularies, standards for data formats, and data handling[17], have so far failed to reach community-wide adoption.

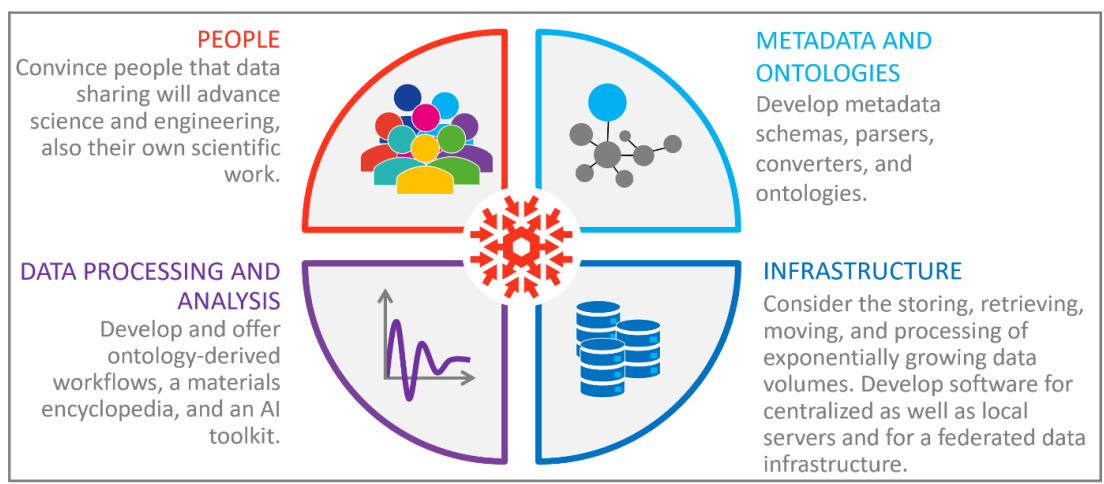

*Figure 1*: Schematic summary of the main challenges addressed by the FAIRmat initiative

*FAIRmat* already started to establish metadata and dictionaries for digital translations of the vocabulary used in different domains. The next step concerns the description of relations between them, hence, the development of ontologies[b]. They will become particularly important when involved workflows are needed. The *NOMAD Metainfo*[18,19,20] stores descriptive and structured information about materials-science data and some interdependencies. Thus, it represents an ontology precursor. There is a lot of discussions within the community, see *e.g.*, Refs.,20, 21, 22, 23, 24, and 25, also concerning the collaboration of *FAIRmat* with *EMMC*[24,25], *OPTIMADE*[23], and NIST[22].

As illustrated in **Error! Reference source not found.**, data-centric materials science requires a complex infrastructure (bottom right). Established standards for data models in materials science will be considered, for example, *CIF (Crystallographic Information Framework)*[26], *CSMD (Core Scientific Metadata Model)*[27], and *NeXus*[28]. Last but not least, acceptance by the researchers requires that the infrastructure also offers support and efficient tools for data processing and analysis (**Error! Reference source not found.**, bottom left).

Other research fields are facing different, yet analogous challenges. International contacts, coordination and collaborations of the various fields are promoted by the GoFAIR[2] initiative, the Research Data Alliance (RDA)[29], the association FAIR-DI[30], CODATA[31], and others. A recent publication by Wittenburg *et al.* on "FAIR Practices in Europe" describes the situation in the areas of humanities, environmental sciences, and natural sciences. While basic concepts and IT tasks are being discussed, true collaborations and reaching the final goal of growing together still need time.

**Preparing the research of tomorrow**

Putting what is sketched above into practice, is a rocky road. To motivate the community to join a culture of extensive data sharing, *FAIRmat*'s policy is to *lead by examples*. Two issues are obviously important in order to speed up the process and trigger active support: (i) Successful, living examples of daily data-centric research[32] to demonstrate that and how things work and (ii) outreach to the wider community, including the education of future scientists and engineers.

To cope with point (i), *FAIRmat* will demonstrate its approach with specific examples from diverse research fields, including battery research, heterogeneous catalysis, optoelectronics, magnetism and spintronics, multi-functional materials, as well as biophysics. In all of this, *FAIRmat* will prove the synergistic interplay of materials synthesis, sample preparation, experiment, as well as theory and computations, and provide a much more comprehensive picture than the single sub-communities can achieve. As such, *FAIRmat* will not only bring together data and tools but, most notably, also *people* who will learn each other's "*language*". In fact, the necessary width of competences goes along with a diversity in the nomenclature, which can hamper the communication as well as the definition of metadata and ontologies. Likewise, electronic lab notebooks (ELNs) must be standardized to allow seamless integration of data into automatic workflows. Dedicated data-analysis and AI tools shall be developed and demonstrated that help identifying the key descriptive physicochemical parameters[33,34,35,36]. This will allow for predictions that go beyond the immediately studied systems and will reveal trends and enable identifying materials with statistically exceptional properties[37]. Combining data from different repositories opens additional opportunities.

Let us exemplify with two emerging classes of materials that the exploitation of an efficient data infrastructure will not only be helpful but simply mandatory for the digitalization of materials research[38]. These examples are high-entropy alloys (HEAs) and metal-organic frameworks (MOFs). For these classes, the sheer number of possible, different materials is so large that conventional approaches will never be able to unleash even a small part of the full potential. For HEAs, a number of $10^9$ possible individual composite materials with distinctly different properties have been estimated[39], with many of them showing, *e.g.*, mechanical properties exceeding by far those of conventional alloys. This huge space of materials further contains HEA oxides, with interesting properties in catalysis and energy storage. In the case of MOFs, the situation is even more dramatic. As a result of the huge diversity of the MOF building blocks, inorganic clusters, and multitopic molecules, the number of compounds is unlimited. Even if one limits the building block weight to that of fullerene ($C_{60}$), synthesizing only one replica of each compound would already need more atoms than available on planet Earth. Using AI to analyze the huge amount of experimental information (data for about 100 thousand MOFs are stored in databases[40]), we will be able identify or to predict MOFs with particular properties dictated by envisioned applications[41], *e.g.*, in optoelectronics[42], biomedicine, or catalysis[43].

Turning to point (ii), *i.e.*, to foster awareness for the importance of a *FAIR* scientific data management and stewardship[1], *FAIRmat* will reach out to the present students of physics, chemistry, materials science, and engineering. We aim at educating a new generation of interdisciplinary researchers, offering classes and lab courses, and introduce new curricula. A necessary requirement is to convince teachers, professors and other decision makers. The *FAIRmat* consortium will initiate and organize focused, crosscutting workshops, *e.g.*, together with colleagues from chemistry and biochemistry, astro- and elementary-particle physics, mathematics, and engineering. Some topics may be general, like ontologies or data-infrastructure, others will be more specific, *e.g.,* particular experimental techniques or specific simulation methods. Hands-on training, schools, and hackathons, as well as regular on-line tutorials will be part of our portfolio. Listening to the needs of small communities or groups will make sure that no one is left behind.

While industry is very interested in the availability of data, the materials encyclopedia, and the AI tools, most of them hesitate to contribute own data. Understandably, a company can only survive if they create products that are better or cheaper than those of their competitors. FAIRmat accepts these worries, for example by allowing for an embargo of uploaded data (see above). The NOMAD *Oasis*[45] (see also below), which is a key element of the federated FAIRmat infrastructure, can also be operated behind industrial firewalls as a stand-alone server with full functionality.

Science is an international, open activity. So, clearly, all the concepts and plans are and will be discussed, coordinated, and implemented together with our colleagues worldwide. In fact, the first FAIR-DI Conference on a *FAIR* Data Infrastructure for Materials Genomics had 539 participants from all over the world[44].

Let us end this section by noting that the individual researchers already profit from the data infrastructure, even if we are at an early stage towards the next level of research. For example, countless CPU hours are being saved because computational results are well documented and accessible and don't need to be repeated. Consequently, human time is saved as well and scientists can concentrate on new studies. Students learn faster as they can access extensive reference data. Error or uncertainty estimates are possible and more robust when using well-documented databases. Additional results not documented in publications are available in the uploaded data. Studies that aimed at a specific target can now be used for a different topic (re-purposing). After receiving a DOI (digital object identifier), uploaded data become citable. This also applies to analytics tools. Although the full potential of FAIRmat will require a bigger community to realize and join, already now the spirit of *Findable and AI Ready* research data has attracted substantial attention.

*FAIR* **data infrastructure for materials science**

*FAIRmat* will build a federated infrastructure of many domain-specific data-repository solutions: as few as possible but as many as needed. In *NOMAD*[9], such individual repositories are called '*Oases*'[45] and support the different users' local, domain-specific, individual needs to acquire, manage, and analyze their data. An *Oasis* is a stand-alone service typically connected to a central server, called the *Portal*; but can be also run independently.

All participating groups or institutions will manage their data using the *FAIRmat* frame, a common compute, management, and storage concept, with a central metadata repository. To enable *4V* data processes, 'federated data with centralized metadata' will be the general principle. Selected data may also be stored centrally, if its functionally is beneficial for users or increases the availability of high-value datasets (see **Error! Reference source not found.**).

The *Portal* will be the gateway for users to access all materials-science data. While popular search engines, *e.g.,* Google, search for phrases in generic and mostly text-based properties (domain *agnostic*), we need to search for *precise* criteria in materials-science-specific metadata with their individual scientific notations and semantics. Thus, *FAIRmat* searches are domain-*aware*.

We will implement a common schema for all *FAIRmat* metadata and data. However, what data properties are available for a given type of data differs from method to method and from domain to domain. There will be subsets of common properties for each sub-domain, and these subsets form a hierarchy. For example, experiments and synthesis share a common notion of material, measurement, or sample. This includes tagging samples with RFID (radio frequency identification) or QR (quick response) code labels that are linked to every dataset acquired from them. On top of this hierarchy, and even outside of the materials-science domain, we will always have Dublin-core-style[46] metadata about *who*, *where*, and *when*.

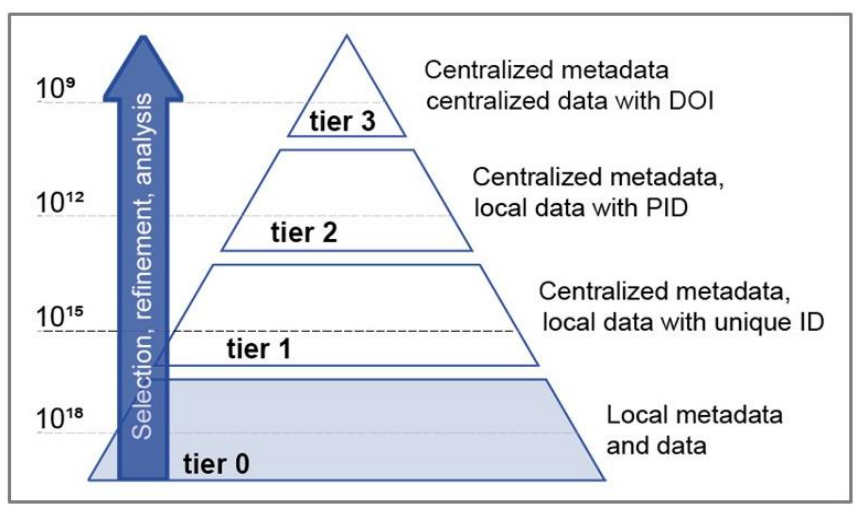

Figure 2: Tiers of metadata and data and their approximate volume in bytes (left axis). As part of the research process, acquired data are filtered, analyzed, and refined. This generally produces smaller and smaller datasets of higher value. Due to the expected data volume, the same level of availability for all data cannot be guaranteed. Therefore, data are categorized into four tiers with different levels of availability, from not shared at all (tier 0) towards published with a DOI (tier 3).

This bottom-up harmonizing of metadata from different sub-domains requires developing data converters and a shared data schema. This will provide more flexibility when connecting many laboratories and new sub-domains than a top-down forced adoption of a new data format.

This hierarchy of common properties will also form the basis for exploring all materials-science data. Similar to an online shop that allows customers to browse different categories of products, with varying criteria depending on the type of product, the central user interface will allow one to browse different sub-domains of materials science based on varying availabilities of data properties. On the top, one may specify general properties, *e.g.*, a material's chemical composition and a scientific method. Then, more criteria will be made available. This way, we will design a common encyclopedia that supports the specific needs of the various material science sub-domains but will also provide more general information to non-experts.

Offering convenient tools for data analysis is an overall goal of FAIRmat. An example is the NOMAD Artificial Intelligence Toolkit[16]. Currently, it provides several Jupyter notebooks, some of them associated with a publication. It is recommended that researches publish their AI analysis as well or modify or advance existing notebooks for their studies. Uploaded notebooks can obtain a DOI so that they are citable. As some data files will be huge, and may be distributed at several servers and cities, the analysis software will use the centralized metadata and extract the needed information from the (huge) data files. For the latter, we will bring the software to the data avoiding transfer of big files.

Other critical issues are long-term and 24/7 data availability (especially in a federated network), safety/security (especially dealing with published vs. unpublished data), data lifecycle (*e.g.*, from raw instrument readings to fully analyzed and publish datasets), linking data between domains, annotating data with a common user identity (*e.g.*, via ORCID[47]), and more.

*FAIR*, reproducible synthesis

Synthesizing materials with well-defined properties in a reproducible fashion is of utmost importance to materials science. Unfortunately, this request is not always fulfilled because it requires controlling a high number of experimental details, and the full entirety of the relevant parameters is typically not known. The concept of data-centric science and the development of AI tools promises to model synthesis more reliably and to identify the relevant set of descriptive parameters and their mutual interdependencies or at least their correlations. Linking synthesis data to data from experimental materials science and theory by common metadata schemas and ontologies will create a new level of the science of material synthesis.

Publicly accessible databases such as Landolt-Börnstein / Springer Materials[48] and the Inorganic Crystal Structure Database[49] contain huge amounts of entries on the properties of crystalline materials, but they lack information on the synthesis. Recently, work based on machine-learning and natural-language-processing techniques started to codify materials synthesis conditions and parameters that are published in journal articles[50,51]. However, typically this information is incomplete, and published information is biased towards reports of successful studies, leaving out failed attempts.

This unsatisfactory situation is rooted in the complexity of the synthesis processes, including elaborate workflows and a large diversity of instruments for characterization. In the realm of *FAIRmat*, we follow the relevant phase transformations occurring during synthesis from the melt, from the gas phase, from solid phases, and from solution. *Synthesis by assembly* complements these classical approaches. The nature of the assembly method is quite different, since collective behavior gives rise to new properties, such as formation of aggregates or self-assembly. **Error! Reference source not found.** depicts the variety of crystal growth methods. Even though Czochralski, Bridgeman, flux growth, and optical floating zone are all melt-growth techniques, *i.e*, they belong to the same type of phase transition, they are distinguished by the contact of the melt to the crucible, the seeding of the single crystal and by thermal gradients, with high impact on crystallinity and impurity content. But even fine details matter. For example, the geometry of the reactors, fluctuations in the impurity content of the source material, the flow of precursors in the reactor, or the miscut and pretreatment of substrates in epitaxial growth may have detrimental effects. At this point, synthesis is often based on experience and tricks, which are not readily shared with others. Obviously, this makes the development of metadata schemas and ontologies a formidable task, and with respect to the *4V*, synthesis struggles mainly with *variety* and *veracity*.

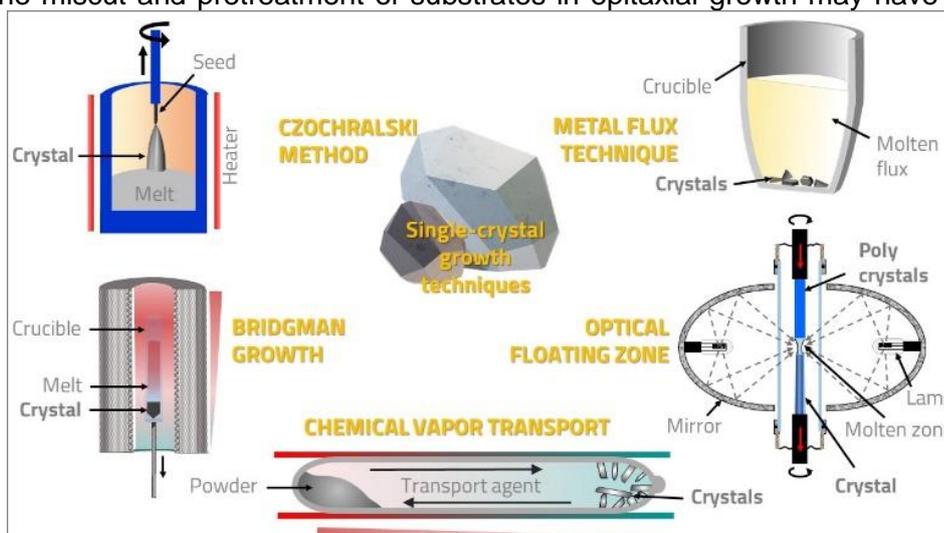

Figure 3: Examples of single crystal growth techniques reflecting a small part of the variety of the synthesis processes. Simplified sketches of the five most common crystal growth methods from the melt (Czochralski, Bridgman, Optical Float Zone, and Metal Flux Growth) and the gas phase (Chemical Vapor Transport).

We started to establish metadata and ontologies following the above-mentioned phase transformations. In order to connect to the other experimental disciplines, *e.g.*, regarding sample characterization, we use a common ELN scheme and laboratory information management systems (LIMS) and uniquely identify the samples as noted in the *infrastructure section*. Thereby, we link the measured physical and chemical properties of a specimen to the synthesis workflow. The ELN and LIMS data are automatically fed into a prototype repository that is currently being developed at the *Leibniz Institute for Crystal Growth*[52].

Once a structured database on synthesis being established, this will allow for computer-aided development of synthesis recipes to fabricate yet unknown materials with tailored properties. Moreover, it will enable comparison of different synthesis methods of the same material in terms of generalized physical and chemical parameters, also linking them to theoretical predictions.

## *FAIR* data in experimental disciplines

Experimental materials science is concerned with the characterization of the atomic and electronic structure of compounds, as well as with determining their electrical, optical, magnetic, thermal, or mechanical properties. Typically, terabytes (sometimes petabytes) of data of one study result in a few plots in a publication. Only a *FAIR* data management of all results, the successful and the failed ones, makes experimental studies reproducible and obviates the necessity to repeat the experiments for a different but related project. In addition, by making all these data available to the community, all will benefit from a statistically more reliable quantification of measurement errors and calibrations.

In experimental materials science, the variety of characterization methods is extremely diverse, and each class of methods has its own equipment and workflows used for generating data. The diversity in vendor-, lab-, instrument-, community-, and operator-specific data formats, presents a substantial challenge with regard to integrate this information into a *FAIR* infrastructure. For the initial period, we concentrate on five experimental techniques (see **Error! Reference source not found.**) with very different frontiers in terms of the *4V* challenges, and largely disjunct and differently structured communities. These are electron microscopy and spectroscopy (EM), angle-resolved photoemission spectroscopy (ARPES), core-level photoemission spectroscopy (XPS), optical spectroscopy (OS), and atom-probe tomography (APT). The amount of generated data

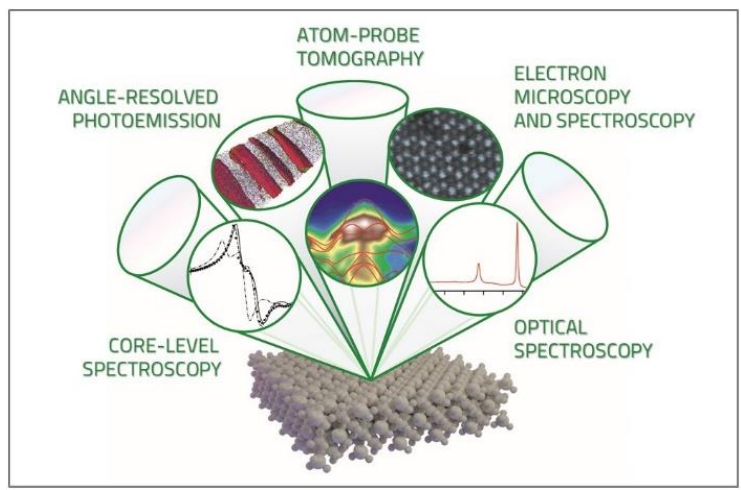

Figure 4: Illustration of five experimental materials characterization techniques chosen to focus on initially, for the purpose of establishing digital *FAIR* data management workflows. Optical spectroscopy (OS), atom-probe tomography (APT), angle-resolved photoelectron spectroscopy (ARPES), electron microscopy (EM), and x-ray photoemission spectroscopy (XPS) are noted explicitly, and the big number of other experimental methods is indicated by the empty 'perspectives'

ranges from a few kB to TB per data set, and also the data rates and data structures differ substantially. With some modern detectors delivering several GBs of data per second, the *volume* and *velocity* challenge is to pre-process, compress, and evaluate/visualize these data. This becomes a more severe *velocity* issue in time-resolved experiments, for which the duration may not even be fixed, but being decided during the observation. Disturbingly, overall, we observe a lack of efficiently and reliably rec ording metadata in a digital form, posing a severe data-*veracity* challenge.

Analogously to establishing FAIR data in synthesis, a strong focus is the customization of inter-operating ELNs and LIMSs, their integration into experimental workflows, and their direct connection to the data repository.

In each of the five selected experimental techniques, activities have commenced to also define domain-specific metadata catalogues and ontologies. In some labs (TEM and spectroscopy), a first, rudimentary protype of an *NOMAD Oasis* has recently been installed with the aim of exploring how it should be further developed towards the requirements of the different sub-domains. Integration of ELNs into the experimental workflow is at different stages of development, ranging from first implementation concepts in APT to a working integration of an ELN database with the data acquisition software in some TEM labs. This also includes tagging of samples with QR labels and automated linking of sample IDs with links to experimental data and time-stamped notes generated by the data acquisition software. Some ARPES groups are presently reorganizing their labs, switching from paper lab books to ELNs. In this context, we note that in a joint effort of different labs, we were able to make vendors of complex equipment to reconsider their previously restrictive and closed data-format policies.

## *FAIR* theory and computations

Materials modeling, in particular including digital twins, is enjoying ever-growing attention thanks to a timely combination of hardware and algorithmic developments[53]. The *NOMAD Laboratory*[9], already implemented a materials-data infrastructure for quantum-mechanical ground-state calculations and *ab initio* molecular dynamics (see the summary in section *data-centric science*, above). Yet, materials modeling also requires force fields and particle-based methods, to capture larger length and longer time scales (see **Error! Reference source not found.**). Implementing such multiscale materials-data infrastructure carries a number of outstanding challenges[54,55]. By considering trajectories, we need to account for both instantaneous and ensemble properties. Also, the heterogeneity of simulation setups, solvers, force fields, but also observables requires an ambitious and coherent strategy to make multiscale modeling *FAIR*. The development of metadata of this field has started only now.

Another crucial task is the response of matter to external stimuli. The physical objects of interest obtained from theory are excitation energies and lifetimes, electronic band gaps, dielectric tensors, various excitations spectra, ionization potentials, all of them having experimental counterparts. The leading methodologies[56] comprise time-dependent DFT (TDDFT), Green-function (GF) techniques, and dynamical mean-field theory (DMFT), implemented in a huge number of different computer codes. The envisaged *FAIRmat* infrastructure will foster the presently often incomplete documentation and facilitate benchmarking and a curation of results.

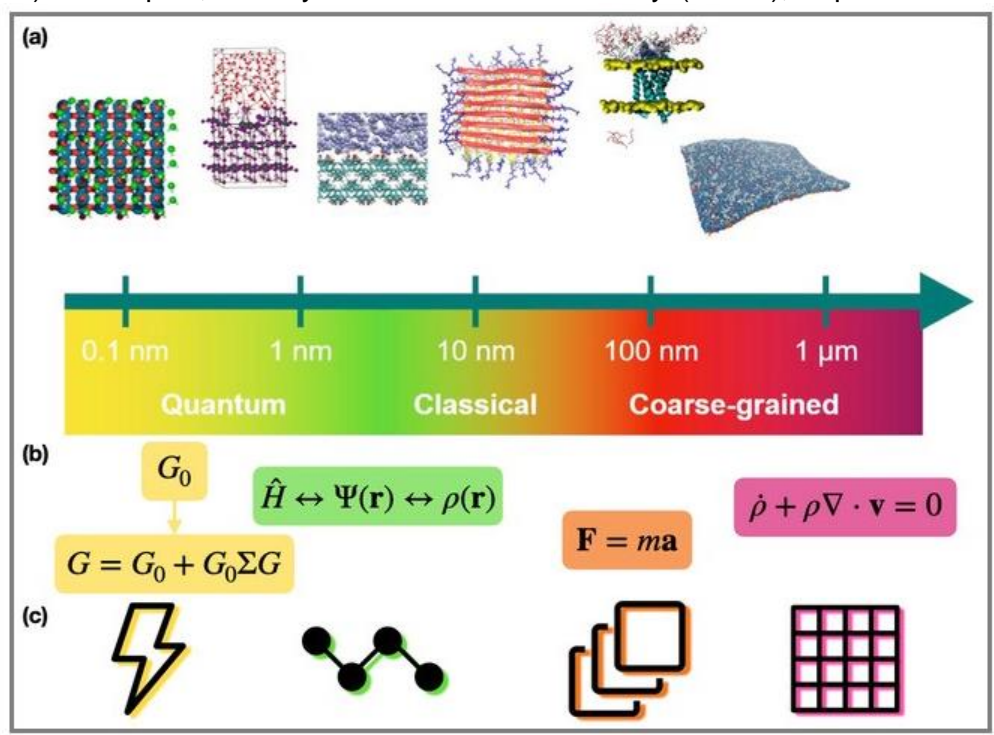

Figure 5: (a) Multiscale challenge of materials simulations, ranging from sub-Ångstrom up to length scales of micrometers and more. (b) Approaches range from excited states, ground-state density functional theory, force-field based molecular dynamics, to continuum. (c) The data infrastructure requires needs to tailor to the different methods, such as trajectories and densities for molecular dynamics and continuum methods, respectively.

Concerning the *4V*, the area of *theory and computations* is severely affected by *variety*, *i.e.*, the heterogeneity of the meaning of the produced data. This refers to the fact that there are many physical equations, even more algorithms, and again more approximations that are implemented in the numerous, very different software packages. While *ab initio* computational materials science has largely assumed a common nomenclature, *e.g.,* for the several hundred exchange-correlation approximations (see, *e.g.*, libxc[57]), this is not yet the case for force fields, dynamical mean field theory, or calculations of fluid dynamics, *etc.*

Somewhat related to the *variety* challenge is *veracity*. Note that we differentiate between *accuracy* and *precision*, where the latter can be checked by comparing results from different software that address the same equations and use the same approximations. While for *ab initio* computational materials science the first important steps have been made[58], for other theoretical approaches such efforts are still missing. Accuracy, in turn, refers to the equations and basic approximations (*e.g.*, the actually used exchange-correlation functional or the force field). Here, error bars are largely missing so far, but for interoperability with experimental results such error estimates need to be developed. Concerning the data *volume*, for molecular dynamics calculations it is hardly possible to store all the information, *i.e.*, the detailed time evolution of the positions of all the atoms (if these are several thousand, as in force field studies) or the electronic charge density (in *ab initio* studies). Here selection and compression strategies will be developed.

**Making the data revolution happen**

Fourteen years after Jim Gray's explication on *Data-Intensive Scientific Discovery*[59], materials science is still dominated by the first three research paradigms that are experiment, theory, and numerical simulations[5]. However, there is now wide consensus that data-centric research and the 4th paradigm, *i.e.*, data mining, new ways of analysis (largely by AI), and visualization, will change if not revolutionize the sciences. Let us stress that the 4th paradigm represents a new way of thinking[5]. It complements but does not replace the previous concepts and approaches. Implementation of this paradigm not only creates new opportunities, but also enhances the traditional approaches through efficient data exchange, better documentation, and a more detailed understanding of what other groups are doing. This will open new horizons for research in basic and engineering sciences, reaching out to industry and society.

So, let us summarize what we need to make the data revolution happen in materials science:

- **Hardware for data storage and handling, advanced analytics, and high-speed networks:** Availability of appropriate hardware is the basic prerequisite to build the described data infrastructure. Then, we also need middleware, *e.g.*, for the efficient exchange of data that are created in or by different digital environments. We add in passing that efficient, near-real-time data analytics will also require advanced hardware as well as soft- and hardware co-design.

- **Development and support of software tools:** New tools are being invented already, for example, for fitting data, removing noise from data, learning rules that are behind patters in data, and identifying "statistically exceptional" data groups[37]. With such rules one will also identify "materials genes", *i.e.*, physical parameters that are related to the processes that trigger, facilitate, or hinder a certain materials property or function. *FAIRmat* will foster the international coordination of such tool developments in the wider materials-science community.

   Another key issue concerns the development of ELNs and LIMS. Such necessary changes of current scientific procedures appear minor if one accepts that it is good scientific practice to document the experimental (or computational) conditions and the results in full detail, so that studies are reproducible. Thereby, data collection (including the comprehensive characterization of the experimental setup) should become as automatic as possible. This sounds like an out-of-date request, but it has not been executed properly, so far, and for data-centric science it is essential. Unfortunately, for some, maybe many studies, an immediate realization is not fully possible, and even the first approximation requires a "phase transition". Owing to the complexity of the field, there is no on-size-fits-all solution.

   All this implies close collaboration between experts from data science, IT infrastructure, software engineering, and the materials-science domain as equal partners. In FAIRmat, this will be realized by a centralized hub of specialists at the Physics Department of the Humboldt University in Berlin.

- **Changing the publication culture and advancing digital libraries:** As noted above, the basic scientific requirement of reproducibility of experimental work is often lacking. This is rooted in the complexity and intricacy of materials synthesis. *FAIRmat* will change this situation. The concept of "clean data", *i.e.*, data that are comprehensively annotated, is being developed (see Ref. 60 and references therein). This is much more elaborate than it sounds, and publications that "just" present and describe such data comprehensively should be appreciated by the community as much as a standard publication in a high-impact journal.

Digital libraries have been and are being built and advanced over the last decade. While there have been ample developments in the fields of life sciences, the situation in materials science is less advanced but improving (*e.g.*, Refs. 61, 62). and in this field metadata catalogues are typically too unspecific to allow for the identification of suitable datasets (*e.g.,* for AI analysis).

The German national research-data infrastructure project[63] (NFDI) promotes all the above discussed points, with the exception of the needed hardware. Although a national effort, it is obviously part of an international activity, and *FAIRmat* has established respective collaborations already. We will support scientists and confirm them in their responsible handling of research data, and we will strive for educating the next generation of researchers and engineers to *actively* engage in order to reach the goals in a good timeframe.

The field is changing, and the community seems mostly convinced about the bearing of this change, but it is still mostly in the role of an observer. If the active scientists don't sign on, the infrastructure will develop without them. Then, in a few years, they need to accept what is there, and it may, unfortunately, not fully serve their needs. The consequences of the whole endeavor may be summarized as follows: The envisaged changes brought by a FAIR data infrastructure will not replace scientists, but scientists who use such infrastructure may replace those who don't.

**Acknowledgement**

This work received funding from the Max Planck Research Network BiGmax, the Humboldt-Universität zu Berlin, and the European Union's Horizon 2020 research and innovation program under the grant agreement Nº 951786, the NOMAD CoE. We thank Sören Auer for his feedback on Digital Libraries, Victoria Coors for her thoughtful work on the figures, and the entire FAIRmat team for shaping the concept and first steps towards its implementation. The project FAIRmat is funded by the Deutsche Forschungsgemeinschaft (DFG, German Research Foundation) – project 460197019.

**Authors contributions:**

C. Draxl (chairperson of FAIRmat) and M. Scheffler (co-chair) created the general concept, lead the writing, and edited the paper on the whole, in close discussion with all other authors and gratefully appreciating comments by referees. They also wrote the *abstract/preface*, and the sections *introduction*, *data-centric materials science*, and *making the data revolution happen*. M. Albrecht and C. Felser (Area A of FAIRmat) took the lead on section *FAIR, reproducible synthesis*; M. Greiner and C.T. Koch on section *FAIR data in experimental disciplines* (Area B of FAIRmat); M. Scheffler., K. Kremer, and T. Bereau on section *FAIR theory and computations* (Area C of FAIRmat), H.J. Bungartz, M. Scheidgen, and W. Nagel on section *FAIR data infrastructure for materials science* (Area D of FAIRmat). C. Draxl, together with C. Wöll and A. Groß (Area E of FAIRmat) and M. Scheffler and M. Aeschlimann (Area F of FAIRmat) created the section *preparing the research of tomorrow*. All authors carefully read and polished the whole paper.

The authors declare no competing interests in relation to the work described.

[1] Wilkinson, M.D., *et al*. The FAIR Guiding Principles for scientific data management and stewardship. *Sci Data* **3**, 160018 (2016).
**This work coined the acronym FAIR that is used now worldwide.**

[2] GoFAIR https://www.go-fair.org/fair-principles/

[3] Materials Genome Initiative (MGI), https://mgi.gov/

[4] FAIRmat - FAIR Data Infrastructure for Condensed-Matter Physics and the Chemical Physics of Solids; a Proposed Consortium of the German Research-Data Infrastructure (NFDI[63]), https://fairdi.eu/fairmat

[5] Draxl, C. and Scheffler, M. Big-Data-Driven Materials Science and its FAIR Data Infrastructure. Perspective in Andreoni, W. and Yip, S. (eds.). *Handbook of Materials Modeling*. Springer, Berlin, 49 (2020).
**Addresses the fourth paradigm of materials research and highlights the related challenges.**

[6] Nature Editorial *Empty rhetoric over data sharing slows science*, 12 June 2017; *Nature* **546**, 327 (2017).

[7] Draxl, C., Illas, F., and Scheffler M. Open data settled in materials theory. *Nature* **548**, 523 (2017); Correspondence to Ref. [6].

[8] Himanen, L., Geurts, A., Foster, A. S., and Rinke, P. Data-Driven Materials Science: Status, Challenges, and Perspectives. *Adv. Sci.* **6**, 1900808 (2019).

[9] The Novel Materials Discovery (NOMAD) Laboratory, https://nomad-lab.eu/
**First materials-science repository fostering sharing and accepting data from the community. See also Ref. 7.**

[10] Draxl, C. and Scheffler, M. The NOMAD Laboratory: From Data Sharing to Artificial Intelligence. *J. Phys. Mater.* **2**, 036001 (2019).

[11] arXiv, https://arxiv.org/, is an open-access repository of electronic preprints approved for posting after moderation, but not peer review. It consists of scientific papers in the fields of mathematics, physics, astronomy, electrical engineering, computer science, quantitative biology, statistics, mathematical finance and economics, which can be accessed online. (Text adopted from Wikipedia).

[12] AFLOW, http://aflow.org, is a globally available database. In March 2021 it contained data of 3,466,057 material compounds with over 679,347,172 calculated properties.

[13] The Materials Project, https://materialsproject.org, provides open web-based access to computed information on known and predicted materials as well as powerful analysis tools to inspire and design novel materials. In March 2021 it contained data of 131,613 inorganic compounds.

[14] OQMD, http://oqmd.org, is a database of DFT calculated thermodynamic and structural properties of 815,654 materials (March 2021).

[15] The NOMAD Encyclopedia allows users to see, compare, explore, and understand computed materials data, https://nomad-lab.eu/encyclopedia

[16] The NOMAD Artificial Intelligence Toolkit contains tools for finding patterns and information in materials science data, https://nomad-lab.eu/AIToolkit
18